\documentclass[%
 aip,
 rsi,
amsmath,amssymb,
reprint,
floatfix
]{revtex4-1}

\usepackage{graphicx}
\usepackage{dcolumn}
\usepackage{bm}

\usepackage[utf8]{inputenc}
\usepackage{mathptmx}
\usepackage{braket}
\usepackage[english]{babel}
\usepackage{xcolor}
\colorlet{RED}{red}
\colorlet{BLUE}{blue}
\usepackage{dcolumn}
\usepackage{bm}
\usepackage[version=4]{mhchem} 
\usepackage{acronym}
\usepackage{adjustbox}
\usepackage{tikz}
\usepackage[toc,page]{appendix}
\usepackage{microtype} 
\usetikzlibrary{calc,shapes.geometric,decorations.pathmorphing,patterns}

\usepackage[margin=2.3cm,bmargin=1cm,footnotesep=1cm]{geometry}

\usepackage[final]{changes}
\definechangesauthor[color=Blue]{NPB}
\usepackage{todonotes}
\setcommentmarkup{\todo[color={authorcolor!20},size=\scriptsize]{#3: #1}}
\newcommand{\note}[2][]{\added[#1,comment={#2}]{}}

\begin{document}


\title{
Coupled Cluster Downfolding Theory: towards efficient many-body algorithms for dimensionality reduction of composite quantum systems
}

\author{Nicholas P. Bauman}
 \affiliation{Physical Sciences and Computational Division, Pacific Northwest National Laboratory, Richland, Washington 99354, United States of America}
\author{Karol Kowalski}
\email{karol.kowalski@pnnl.gov}
 \affiliation{Physical Sciences and Computational Division, Pacific Northwest National Laboratory, Richland, Washington 99354, United States of America}

\date{\today}

\begin{abstract}
  The recently introduced coupled cluster (CC) downfolding techniques for \replaced{reducing the dimensionality}{dimensionality reduction} of quantum many-body problem\added{s} recast the CC formalism in the form of the renormalization procedure allowing\added{,} for the construction of effective (or downfolded)  \replaced{Hamiltonians}{Hamiltionians} in small-dimensionality sub-space, usually identified with the so-called active space, of the entire Hilbert space. 
  The resulting downfolded Hamiltonians \replaced{integrate}{intergrate} out the external (out-of-active-space) Fermionic degrees of freedom from the internal \deleted{ones} (in-the-active-space) parameters of the wave function\added{,} which can be determined as components of the eigenvectors of the  downfolded Hamiltonians in the active space. \replaced{This paper will}{In this paper we will} discuss the extension of  non-Hermitian (associated with standard CC formulations) and Hermitian (associated with the unitary CC approaches) downfolding formulations to composite quantum systems. \replaced{The non-Hermitian formulation can provide a platform for developing local CC approaches, while the Hermitian one can serve as an ideal foundation}{While the non-Hermitian one can provide a development platform for the development of local CC formulations the former one can serve as an ideal environment} for developing various quantum computing applications based on the limited quantum resources. We also discuss the algorithm for extracting \added{the} semi-analytical form of the inter-electron interactions in the active spaces. 
\end{abstract}
\maketitle

\section{Introduction}
The coupled cluster (CC) methodology \cite{coester58_421,coester60_477,cizek66_4256,paldus72_50,purvis82_1910,jorgensen90_3333,paldus07,crawford2000introduction,bartlett_rmp}  is a driving engine of high-precision simulations in physics, chemistry, and material sciences. Several properties of CC made it especially efficient in capturing correlation effects in many-body quantum systems ranging from quantum field theory, 
\cite{funke1987approaching,kummel2001post,hasberg1986coupled,bishop2006towards,ligterink1998coupled}
quantum hydrodynamics,
\cite{arponen1988towards,bishop1989quantum}
\added{and} nuclear structure theory 
\cite{PhysRevC.69.054320,PhysRevLett.92.132501,PhysRevLett.101.092502}
to quantum chemistry \cite{scheiner1987analytic,sinnokrot2002estimates,slipchenko2002singlet,tajti2004heat,crawford2006ab,parkhill2009perfect,riplinger2013efficient,yuwono2020quantum} 
and material sciences.
\cite{stoll1992correlation,hirata2004coupled,katagiri2005equation,booth2013towards,degroote2016polynomial,mcclain2017gaussian,wang2020excitons,PhysRevX.10.041043}  
In this article\added{,} we will mainly focus our attention on the application to quantum chemistry. 
Many appealing features of the single-reference (SR) CC
formalism (which will be the main focus of the present discussion)  in applications to chemical systems originate in the exponential parametrization of the ground-state wave function and closely related linked cluster theorem.\cite{brandow67_771,lindgren12,shavitt2009many} The last feature assures \replaced{the so-called}{the-so-called} additive separability of the calculated energies in the non-interacting sub-system limit, which plays \added{a} critical role in \added{the} proper description of various chemical transformations such as chemical reactions that include bond breaking and \replaced{bond-forming}{bond forming} processes. The linked cluster theorem \added{also} plays \deleted{also} a crucial role in \replaced{designing}{the design of} formalisms that can provide the \deleted{so-called} chemical accuracy needed for \deleted{accurate} predicting spectroscopic data, reaction rates, and thermochemistry data. The \replaced{best-known}{best known} example of such class of methods is the CCSD(T) formalism,\cite{raghavachari89_479} which combines the iterative character of the CCSD formalism \cite{purvis82_1910}  (CC with single and double excitations) with perturbative  techniques for  determining CC energy corrections due to connected triple excitations. Over the last few decades\added{,} the CCSD(T)-type formulations have been  refined to provide accurate description of bond-breaking processes. Among several formulations that made it possible \deleted{to happen} were\replaced{ the}{:}  method of moments of coupled cluster equations and renormalized approaches,\cite{bookmmcc,mmcc1,mmcc2,crcc,cu2o2,crccopen,crccrev,deustua2017converging,deustua2018communication,bauman1}
perturbative formulations based on the $\Lambda$-operator (defining the left eigenvectors of the similarity transformed Hamiltonians)
\cite{stanton1_t,stanton2_t,crawford_t,stan1,stan2,gwal1,gwal3,sohir1,ybom1}
and other techniques.\cite{tqf,mebar1,robkn,ugur1}
One should also mention \added{the} tremendous effort in formulating reduced-scaling or local formulations of the CC methods to extend the applicability of the CC formalism across spatial scales.\cite{hampel1996local,schutz2000low,schutz2000local,schutz2001low,li2002linear,li2006efficient,li2009local,li2010multilevel,neese2009efficient,neese2009accurate,Neese16_024109,riplinger2013natural,pavosevic2016}\note{added references} In several cases, the extension of local formulations was possible for linear response CC theory \cite{d2020pno++} and excited state CC formulations based on the equation-of-motion formalism.
\cite{dutta2016speeding,peng2018state}

Recently, \deleted{an} interesting aspects of SR-CC were discussed using \added{the} sub-system embedding sub-algebras (SES) approach\added{,}\cite{safkk,kowalski2021dimensionality}  where we demonstrated  that the CC energy can be calculated in \added{an} alternative way to \added{the} standard CC energy formula. Instead of using standard energy expression, one can obtain the same energy by diagonalizing the downfolded/effective \replaced{Hamiltonian}{Hamitlonian} in method-specific active space(s) generated by appropriate sub-system embedding sub-algebras. The SR-CC theory provide\added{s} a rigorous algorithm \added{for} how to construct these Hamiltonians using the external, with respect to the active space, class of cluster amplitudes.\cite{safkk}
\replaced{Shortly after this discovery, these}{These} results for static SR-CC formulations \replaced{were}{have been shortly after} extended to the time domain.\cite{downfolding2020t}\note{added a reference here} Following similar concepts as in the static case and assuming that the external time-dependent cluster amplitudes are known or can be effectively approximated,  it was shown that the quantum evolution of the entire system can be generated \added{in the active space} by time-dependent downfolded Hamiltonian\deleted{ in the active space}. 
Another interesting aspect of CC SES downfolding is the possibility of integrating several SES CC eigenvalue problems corresponding to various active spaces into a computational flow or quantum flow as discussed in Ref.\cite{kowalski2021dimensionality}, where we demonstrated that the flow equations are fully equivalent to the standard approximations given by cluster operators defined by unique internal excitations involved in the active-space problems defining the flow. This feature provides a natural language for expressing the sparsity of the system. In contrast to other local CC approaches, the CC quantum flow equations can effectively embrace the concept of localized orbital pair\added{s} at the level of effective Hamiltonian acting in \added{the} appropriate  active space. 

The SES CC downfolded Hamiltonians are non-Hermitian operators, which limits their utilization in quantum computing. Instead, using double unitary CC (DUCC) Ansatz,\cite{bauman2019downfolding,bauman2019quantumex,downfolding2020t,kowalski2021dimensionality} one can derive the \added{active-space} many-body form of Hermitian downfolded Hamiltonians\deleted{ in active spaces}. In contrast to the SR-CC, the DUCC-based effective Hamiltonians are expressed in terms of non-terminating expansions  involving anti-Hermitian cluster operators defined by external type excitations/de-excitations. Several approximate forms of DUCC Hamiltonians have been tested in the context of quantum simulations, showing \added{the} potential of DUCC downfolding in reproducing exact ground-state energy in small active spaces.\cite{bauman2019downfolding,bauman2020variational} In particular, the downfolded Hamiltonians have been integrated with various quantum solvers\added{,} including Variational Quantum Eigensolvers (VQE)
\cite{peruzzo2014variational,mcclean2016theory,romero2018strategies,PhysRevA.95.020501,Kandala2017,kandala2018extending,PhysRevX.8.011021,huggins2020non,cao2019quantum} 
and Quantum Phase Estimation (QPE)\added{,}\cite{Kitaev:97,nielsen2002quantum,luis1996optimum, cleve1998quantum,berry2007efficient,childs2010relationship,seeley2012bravyi,wecker2015progress,haner2016high,poulin2017fast}
to calculate ground-state potential energy surfaces corresponding to breaking \added{a} single chemical bond.  

In this paper\added{,} we will briefly review the current status of the downfolding methods and 
\deleted{will} provide  further extension of the CC downfolding methods to multi-component system\added{s}. As a specific example, we choose a composite system defined by Fermions of \deleted{the} type A and Fermions of \deleted{the} type B. This is a typical situation encountered for certain classes of non-Born-Oppenheimer dynamics and nuclear structure theory. 
The discussed formalism can be easily extend\added{ed} to other types of systems composed of Fermions and Bosons as encountered in the descriptions of \replaced{polaritonic}{poliratonic} systems. We believe\deleted{,} that these formulations will pave the way for more realistic quantum simulations of multi-component systems.

\section{CC theory} 
The SR-CC theory utilizes the exponential representation of the ground-state wave function 
$|\Psi\rangle$\replaced{,}{:} 
\begin{equation}
|\Psi\rangle = e^T |\Phi\rangle \;,
\label{cc1}
\end{equation}
where $T$ and $|\Phi\rangle$ represent the so-called cluster operator and single-determinantal reference function. The cluster operator can be represented through its many-body components 
$T_k$
\begin{equation}
    T = \sum_{i=1}^M T_k \;,
    \label{cc2}
\end{equation}\note{added a comma at the end of equation}
where individual component $T_k$ takes the form\deleted{:} 
\begin{equation}
T_k = \frac{1}{(k!)^2} \sum_{i_1,\ldots,i_k; a_1,\ldots, a_k} t^{i_1\ldots i_k}_{a_1\ldots a_k} E^{a_1\ldots a_k}_{i_1\ldots i_k} \;,
\label{xex}
\end{equation}\note{added commas in the list of indicies in the sum}
where indices $i_1,i_2,\ldots$ ($a_1,a_2,\ldots$) refer to occupied (unoccupied) spin orbitals in the reference function $|\Phi\rangle$.
The excitation operators $E^{a_1\ldots a_k}_{i_1\ldots i_k} $ are defined through strings of standard creation ($a_p^{\dagger}$) and annihilation ($a_p$)
operators
\begin{equation}
E^{a_1\ldots a_k}_{i_1\ldots i_k}  = a_{a_1}^{\dagger}\ldots a_{a_k}^{\dagger} a_{i_k}\ldots a_{i_1} \;,
\label{estring}
\end{equation}
where creation and annihilation operators satisfy the following anti-commutation rules\added{:} 
\begin{equation}
[a_p,a_q]_+ =
[a_p^{\dagger},a_q^{\dagger}]_+ = 0    \;, \label{comm1}
\end{equation}
\begin{equation}
[a_p,a_q^{\dagger}]_+ = \delta_{pq} \;.\label{comm2}
\end{equation}
When \deleted{the} $M$ in \added{the} summation in Eq.\added{ }(\ref{cc2}) is equal to the number of correlated electron ($N_e$) then the corresponding CC formalism is equivalent to the FCI method, otherwise for \replaced{$M<N_e$}{$N<N_e$} one deals with the standard approximation schemes. Typical CC formulations such as CCSD, CCSDT, and CCSDTQ correspond to $M=2$, $M=3$, and $M=4$ cases, respectively. 

The equations 
for cluster amplitudes $t^{i_1\ldots i_k}_{a_1\ldots a_k}$ and ground-state energy $E$ can be obtained by introducing Ansatz (\ref{cc1}) into the Schr\"odinger equation and projecting onto $P+Q$ space, where $P$ and $Q$ are the projection operator onto the reference function and \replaced{the}{a} space of excited Slater determinant\added{s} obtained by acting with the cluster operator onto the reference function $|\Phi\rangle$, i.e., 
\begin{equation}
    (P+Q) H e^T |\Phi\rangle = E (P+Q) e^T |\Phi\rangle \;,
    \label{sch1}
\end{equation}
where $H$ \replaced{represents the}{represent} electronic Hamiltonian. The above equation is the so-called \replaced{energy-dependent}{energy dependent} form of the CC equations, which corresponds to the eigenvalue problem only in \added{the} exact wave function limit 
when $T$ contains all possible excitations. \replaced{A}{For a}pproximate CC formulations \replaced{{\it do not}}{{\it does not}} represent the eigenvalue problem. At the solution, the energy-dependent CC equations are equivalent to the 
\replaced{energy-independent}{energy independent} or connected form of the CC equations:
\begin{eqnarray}
Qe^{-T}He^T |\Phi\rangle &=& 0 \;, \label{coeq1} \\
 \langle\Phi|e^{-T} H e^{T} |\Phi\rangle &=& E\;. \label{coeq2}
\end{eqnarray}
Using Baker-Campbell-Hausdorff (BCH) formula and Wick's theorem one can show that only \added{connected} diagrams contribute to Eqs.(\ref{coeq1}) and (\ref{coeq2}). For \deleted{the} notational convenience\added{,} one often uses the 
\deleted{so-called} similarity transformed Hamiltonian $\bar{H}$, defined as
\begin{equation}
\bar{H} = e^{-T} H e^T \;. \label{hbar1}
\end{equation}

\section{Non-Hermitian CC downfolding}
The main idea of SR-CC non-Hermitian downfolding hinges upon the 
characterization of sub-systems of a quantum system of interest in terms of active spaces or commutative sub-algebras of excitations that define corresponding active space. This is achieved by introducing  sub-algebras of algebra $\mathfrak{g}^{(N)}$ generated by   
$E^{a_l}_{i_l}=a_{a_l}^{\dagger} a_{i_l}$ operators in the particle-hole  representation defined  with respect to the reference $|\Phi\rangle$. As a consequence of using \added{the} particle-hole formalism\added{,}  all generators 
commute, i.e., $[E^{a_l}_{i_l},E^{a_k}_{i_k}]=0$\added{,} and algebra $\mathfrak{g}^{(N)}$  (along with all sub-algebras considered here) is commutative.
The CC SES approach utilizes  class of sub-algebras  of commutative
$\mathfrak{g}^{(N)}$ algebra,  which contain all possible excitations
$E^{a_1\ldots a_m}_{i_1\ldots i_m}$ needed to generate  all possible excitations from a subset of active occupied orbitals (denoted as $R$\added{, $\lbrace R_i, \; i=1,\ldots,x_R \rbrace$})
to a subset of active virtual orbitals (denoted as $S$\added{, $\lbrace S_i, \; i=1,\ldots,y_s \rbrace$}) defining active space. 
These sub-algebras will be designated as $\mathfrak{g}^{(N)}(R,S)$.
\deleted{In the following discussion, we will use $R$ 
and $S$ 
notation for subsets of occupied and virtual active orbitals $\lbrace R_i, \; i=1,\ldots,x \rbrace$ and 
$\lbrace S_i, \; i=1,\ldots,y \rbrace$, respectively} \replaced{S}{(s}ometimes it is convenient to use alternative notation
$\mathfrak{g}^{(N)}(x_R,y_S)$ where numbers of active orbitals in $R$ and $S$ orbital  sets, \replaced{$x_R$ and $y_S$}{$x$ and $y$}, respectively,  are explicitly called out\replaced{.}{).} 
\deleted{Of special interest in building various approximations are sub-algebras that include all $n_v$ virtual orbitals \replaced{($y_S=n_v$)}{($y=n_v$)} - these sub-algebras will be denoted as 
$\mathfrak{g}^{(N)}(x_R)$.}
As discussed in \replaced{Ref.\onlinecite{safkk},}{Ref.\cite{safkk}} configurations  generated by elements of $\mathfrak{g}^{(N)}(x_R,y_S)$\added{,}  along with the reference function\added{,}
span the complete active space (CAS) referenced to as the CAS($R,S$)
(or \replaced{equivalently}{eqivalently} CAS($\mathfrak{g}^{(N)}(x_R,y_S)$)).

In Refs.\added{ }\onlinecite{safkk,downfolding2020t,kowalski2021dimensionality}\added{,} we explored the effect of partitioning of the cluster operator induced by general  sub-algebra $\mathfrak{h}=\mathfrak{g}^{(N)}(x_R,y_S)$, where \added{the} cluster operator $T$\replaced{, given by Eq. (\ref{cc2}), is represented as}
{(\ref{cc2})}\note{The word 'equation' was misspelled and therefore the equation was inline rather than a separate section.}
\begin{equation}
T=T_{\rm int}(\mathfrak{h})+T_{\rm ext}(\mathfrak{h}) \;,
\label{deco1}
\end{equation}
where $T_{\rm int}(\mathfrak{h})$ belongs to $\mathfrak{h}$ while 
$T_{\rm ext}(\mathfrak{h})$ does no belong to $\mathfrak{h}$.
If the expansion \replaced{$T_{\rm int}|\Phi\rangle$ }{$T_{\rm int}\Phi\rangle$} produces all Slater determinants (of the same symmetry as the $|\Phi\rangle$ state) in the active space, we call $\mathfrak{h}$ \added{the} {\it sub-system embedding sub-algebra} for \added{the} CC formulation defined by the $T$ operator. 
In \replaced{Ref. \onlinecite{safkk},}{Ref.\cite{safkk}} we showed that each CC approximation has its own class \added{of} SESs. 

A direct consequence of existence of the  SESs for standard CC approximations is the fact that the corresponding energy can be calculated, in an alternative way to \replaced{Eq. (\ref{coeq2})}{Eq.()}, as an eigenvalue of the active-space non-Hermitian eigenproblem
 \begin{equation}
    H^{\rm eff}(\mathfrak{h})
    e^{T_{\rm int}(\mathfrak{h})}|\Phi\rangle = 
    E e^{T_{\rm int}(\mathfrak{h})}|\Phi\rangle  \;.
\label{seseqh}
\end{equation}
where 
\begin{equation}
H^{\rm eff}(\mathfrak{h})=(P+Q_{\rm int}(\mathfrak{h})) \bar{H}_{\rm ext}(\mathfrak{h}) (P+Q_{\rm int}(\mathfrak{h}))\;
\label{heffses}
\end{equation}
and 
\begin{equation}
\bar{H}_{\rm ext}(\mathfrak{h})=e^{-T_{\rm ext}(\mathfrak{h})} H e^{T_{\rm ext}(\mathfrak{h})} \;.
\label{heffdef}
\end{equation}
In Eq.(\ref{heffses}) the projection operator $Q_{\rm int}(\mathfrak{h})$
\deleted{is a projection operators onto all excited Slater determinants with respect to the reference function $|\Phi\rangle$. Equivalently, based on the definition of SES, the $Q_{\rm int}$} is a projection operator on a sub-space spanned by  all Slater determinants generated by $T_{\rm int}(\mathfrak{h})$ \deleted{when} acting onto $|\Phi\rangle$. 

Since in the definition of the effective Hamiltonian, \replaced{Eqs. (\ref{heffses}) and (\ref{heffdef})}{Eqs.(\ref{heffses},\ref{heffdef})}, only $T_{\rm ext}(\mathfrak{h})$ is involved, one can view \added{the} SES CC formalism with the resulting active-space eigenvalue problem\replaced{, Eq. (\ref{seseqh}),}{(\ref{seseqh})} as a specific form of renormalization procedure where external parameters defining the corresponding wave function are integrated out. One should also mention that calculating \added{the} CC energy as an eigenvalue problems, \replaced{as described by Eq. (\ref{seseqh})}{Eq.(\ref{seseqh})}, is valid for {\it any SES for a given CC approximation given by cluster operator $T$}.
According to this general result,  the standard CC energy expression\replaced{, shown by Eq. (\ref{coeq2}),}{
(\ref{coeq2})} can be reproduced when one uses trivial sub-algebra, which contains no excitations (i.e., active space \replaced{contains}{contiains} $|\Phi\rangle$ only).

The existence of alternative ways of calculating CC energy opens alternative ways of constructing new classes of approximations. For example, if one integrates several eigenvalues problems corresponding 
to SESs $\mathfrak{h}_i,\;(i=1,\ldots,M)$ into a quantum flow equations (QFE) discussed in \replaced{Ref. \onlinecite{kowalski2021dimensionality}}{Ref.\cite{kowalski2021dimensionality}}, i.e.,\deleted{:}\note{edited punctuation in the equation}
\begin{widetext}
\begin{equation}
    H^{\rm eff}(\mathfrak{h}_i)
    e^{T_{\rm int}(\mathfrak{h}_i)}|\Phi\rangle = 
    E e^{T_{\rm int}(\mathfrak{h}_i)}|\Phi\rangle  \;
    (i=1,\ldots,M)\;.
\label{seseqhf}
\end{equation}
\end{widetext}
In \replaced{Ref. \onlinecite{safkk}}{Ref.\onlinecite{safkk}} we demonstrated that at the solution, the solution of the QFE is equivalent to the solution of standard CC equations in the form \replaced{of Eqs. (\ref{coeq1}) and (\ref{coeq2}) }{(\ref{coeq1},\ref{coeq2})} defined by cluster operator $T$ which is a combination of all {\em unique} excitations included in 
$T_{\rm int}(\mathfrak{h}_i)$ $(i=1,\ldots,M)$ operators. This is can be symbolically expressed as 
\begin{equation}
    T= \bigcup_{i=1}^{M} T_{\rm int}(\mathfrak{h}_i)\;.
    \label{app3bb}
\end{equation}
These two equivalent representations allows one also to form the following important corollary: \\
%
\noindent {\bf Corollary (or the equivalence theorm)}
{\em For certain forms of cluster operator $T$\added{,} the standard connected form of the CC equations given by \replaced{Eqs. (\ref{coeq1}) and (\ref{coeq2}) }{Eqs.(\ref{coeq1},\ref{coeq2})} can be replaced by quantum flow equations composed of non-Hermitian eigenvalue problems\added{, Eq. }(\ref{seseqhf}).
}  \\
The above corollary play\added{s} an important role in defining reduced-scaling formulations. This is a consequence of the fact that each sub-problem 
corresponding to sub-algebra $\mathfrak{h}_i$ has the associated form of the effective Hamiltonian $H^{\rm eff}(\mathfrak{h}_i)$, which allows to define one body-density matrix for the sub-system and select the sub-set of the most important cluster  amplitudes  in the $T_{\rm int}(\mathfrak{h}_i)$ operator. For example, when \added{a} localized orbital basis set is used\added{,} this procedure can be used to define the so-called orbital pairs at the level of the effective Hamiltonian, which is a \replaced{significant}{big} advantage compared to the existing local CC approaches. This procedure can \replaced{also be}{be also} extended to other systems driven by  different types of interactions such as \deleted{encountered} in nuclear structure theory or quantum lattice models, 
where the extension of the standard local CC formulations as used in quantum chemistry is not obvious.

\section{Hermitian CC Downfolding}

In order to employ downfolding methods  in quantum simulations, one has to find \replaced{a}{an alternative}  way to construct Hermitian effective Hamiltonians. 
This goal can be achieved by employing the double unitary coupled Ansatz (DUCC),\cite{downfolding2020t} where the ground-state wave function is represented as 
\begin{equation}
        |\Psi\rangle=e^{\sigma_{\rm ext}(\mathfrak{h})} e^{\sigma_{\rm int}(\mathfrak{h})}|\Phi\rangle \;,
\label{ducc1}
\end{equation}
where $\sigma_{\rm ext}(\mathfrak{h})$ and $\sigma_{\rm int}(\mathfrak{h})$ are general-type anti-Hermitian operators
\begin{eqnarray}
\sigma_{\rm int}^{\dagger}(\mathfrak{h}) &=&  -\sigma_{\rm int}(\mathfrak{h}) \;, \label{sintah} \\
\sigma_{\rm ext}^{\dagger}(\mathfrak{h}) &=&  -\sigma_{\rm ext}(\mathfrak{h}) \;. \label{sintah2}
\end{eqnarray} 
In analogy to the SR-CC case, all cluster amplitudes defining $\sigma_{\rm int}$ cluster operator carry active indices only (or indices of active orbitals defining given $\mathfrak{h}$). The external part $\sigma_{\rm ext}(\mathfrak{h})$ is defined by amplitudes carrying at least one inactive orbital index. However, in contrast to the SR-CC approach,  internal/external parts of anti-Hermitian cluster operators are not defined in terms of excitations belonging explicitly to a given sub-algebra\added{,} but rather by indices defining active/inactive orbitals specific to a given $\mathfrak{h}$.
Therefore $\mathfrak{h}$ will be used here in the context of CAS's generator. Another difference with the SR-CC downfolding lies  in the fact that while for the SR-CC cases components of cluster operators 
$T_{\rm int}(\mathfrak{h})$ and $T_{\rm ext}(\mathfrak{h})$ were commuting 
as a consequence of particle-hole formalism employed, in the unitary case, 
the operators forming $\sigma_{\rm int}(\mathfrak{h})$ and 
$\sigma_{\rm ext}(\mathfrak{h})$ are non-commuting.

Employing DUCC Ansatz\replaced{, Eq. (\ref{ducc1}),}{ (\ref{ducc1})} one can show that in analogy to the SR-CC case, the energy of the entire system (once the exact form of $\sigma_{\rm ext}(\mathfrak{h})$ operator is known) can be calculated through the diagonalization of the effective/downfolded Hamiltonian in SES-generated active space, i.e., 
\begin{equation}
        H^{\rm eff}(\mathfrak{h}) e^{\sigma_{\rm int}(\mathfrak{h})} |\Phi\rangle = E e^{\sigma_{\rm int}(\mathfrak{h})}|\Phi\rangle,
\label{duccstep2}
\end{equation}
where
\begin{equation}
        H^{\rm eff}(\mathfrak{h}) = (P+Q_{\rm int}(\mathfrak{h})) \bar{H}_{\rm ext}(\mathfrak{h}) (P+Q_{\rm int}(\mathfrak{h}))
\label{equivducc}
\end{equation}
and 
\begin{equation}
        \bar{H}_{\rm ext}(\mathfrak{h}) =e^{-\sigma_{\rm ext}(\mathfrak{h})}H e^{\sigma_{\rm ext}(\mathfrak{h})}.
\label{duccexth}
\end{equation}
The above results means that when the external cluster amplitudes are known (or can be effectively approximated), in analogy to single-reference SES-CC formalism, the energy (or its approximation) can be calculated by diagonalizing \added{the} Hermitian effective/downfolded Hamiltonian\replaced{, given by Eq. (\ref{equivducc}),}{(\ref{equivducc}))} in the active space using various quantum or classical diagonalizers. 

The analysis of the many-body structure of the 
$\sigma_{\rm int}(\mathfrak{h})$ and 
$\sigma_{\rm ext}(\mathfrak{h})$ operators \cite{downfolding2020t} shows that they can be approximated in a unitary CC manner:
\begin{eqnarray}
\sigma_{\rm int}(\mathfrak{h}) &\simeq& T_{\rm int}(\mathfrak{h}) - T_{\rm int}(\mathfrak{h})^{\dagger} \;, \label{sint} \\
\sigma_{\rm ext}(\mathfrak{h}) &\simeq& T_{\rm ext}(\mathfrak{h}) - T_{\rm ext}(\mathfrak{h})^{\dagger} \;, \label{sext}
\end{eqnarray}
where $T_{\rm int}(\mathfrak{h})$ and $T_{\rm ext}(\mathfrak{h})$ are SR-CC-type internal and external cluster operators. 

To make a practical use of Hermitian downfolded Hamiltonian\added{s, Eq.} (\ref{duccstep2})\added{,} in quantum calculations one has to deal with  non-terminating expansions \added{of Eq.} (\ref{duccexth}) and determine approximate form of the  $T_{\rm ext}(\mathfrak{h})$ operator to approximate its anti-Hermitian counterpart
$\sigma_{\rm ext}(\mathfrak{h})$ according to \replaced{Eq. (\ref{sext})}{Eqs.(\ref{sint},\ref{sext})}.
In recent studies\added{,} we demonstrated the feasibility of approximations based on the finite commutator expansion. We also demonstrated that 
$T_{\rm ext}(\mathfrak{h})$\added{,} provided by the CCSD formalism\added{,} can efficiently be used in building approximate form of the downfolded Hamiltonians. 
In particular, in this paper we will consider two approximate representations of the downfolded Hamiltonians (A and B) defined by the following expressions for $\bar{H}_{\rm ext}(\mathfrak{h})$:
\begin{widetext}
\begin{eqnarray}
\bar{H}_{\rm ext}^{(A)} &=& H+[H,\sigma_{\rm ext}(\mathfrak{h})] +\frac{1}{2}[[F_N,\sigma_{\rm ext}(\mathfrak{h})],\sigma_{\rm ext}(\mathfrak{h})] \;,
\label{comme1} \\
\bar{H}_{\rm ext}^{(B)} &=& H+[H,\sigma_{\rm ext}(\mathfrak{h})]+\frac{1}{2}[[H,\sigma_{\rm ext}(\mathfrak{h})],\sigma_{\rm ext}(\mathfrak{h})]+\frac{1}{6} [[[F_N,\sigma_{\rm ext}(\mathfrak{h})],\sigma_{\rm ext}(\mathfrak{h})],\sigma_{\rm ext}(\mathfrak{h})] \;, \label{comme2}
\end{eqnarray}
\end{widetext}
where $F_N$-dependent commutators were introduced to provide perturbative consistency of single- (C1) and double-commutator (C2) expansions.

As a numerical example illustrating the efficiency of approximations C1 and C2
we use the LiF molecule at $1.0R_e$, $2.0R_e$, and $5.0R_e$
Li-F distances where \replaced{$R_e=1.5639\text{ \AA}$}{$R_e=1.5639\AA$}. All calculations were performed using \added{the} cc-pVTZ basis set 
\cite{dunning89_1007} (employing spherical representation of $d$ orbitals). The calculations using downfolded Hamiltonians C1 and C2 were performed  employing  restricted Hartree-Fock (RHF) orbitals and active spaces composed of 13 lowest\added{-}lying orbitals (6 occupied and 7 virtual). The results of the diagonalization of the downfolded Hamiltonians are shown in Table \ref{table1}. The C1 and C2 energies are compared with the CCSD, CCSDT, and CCSDT(2)$_Q$ \cite{hirata2004combined} energies obtained with all orbitals correlated and the CCSDTQ formalism in \added{the} active space, which represent nearly exact diagonalization of the electronic Hamiltonian in \added{the} active space. 

A comparison of the RHF and CCSDTQ-in-active-space results indicates that the active space used reproduces only a very small part of the total correlation energy approximately represented by the CCSDT(2)$_Q$ results.  In spite of this deficiency in the active space choice, the C2 DUCC approximation yields 9.99, 19.70, and 4.53 milliHartree of error with respect to the CCSDT(2)$_Q$ energies for 1.0R$_e$, 2.0R$_e$, and 5.0R$_e$ geometries, respectively. These errors should be collated with the errors of \added{the} CCSDTQ-in-active-space approach of 310.97, 311.20, and 299.49 milliHartree. As seen from Table \ref{table1}\added{,} the inclusion of double commutator (C2 approximation) results in a significant improvements of the energies obtained with the C1 scheme.

\renewcommand{\tabcolsep}{0.2cm}
\begin{center}
  \begin{table*}
    \centering
    \caption{A comparison of the CC energies obtained for the LiF model in the cc-pVTZ basis set (see text for more details) with C1 and C2 energies obtained in \added{the} active space defined by \added{the} 13 lowest\added{-}lying RHF orbitals.}
    \begin{tabular}{lccc} \hline \hline  \\
  Method &  1.0R$_e$ &  2.0R$_e$ & 5.0R$_e$  \\
\hline \\
RHF  & -106.980121 & -106.850430 & -106.728681 \\[0.1cm]
CCSD & -107.283398 & -107.153375 & -107.022451 \\[0.1cm]
CCSDT   &  -107.291248  &  -107.161817 & -107.028098 \\[0.1cm]
CCSDT(2)$_Q$ & -107.291453 & -107.162103 & -107.028288 \\[0.1cm]
CCSDTQ  & -106.980480 & -106.850899 &  -106.728(8)  \\[-0.1cm]
in act. space &   &    &                            \\[0.1cm]
C1   & -107.276752  & -107.147287   &  -107.019105 \\[0.1cm]
C2   & -107.281461	 & -107.142401   &  -107.032819 \\[0.2cm]
    \hline \hline 
    \end{tabular}
    \label{table1}
  \end{table*}
\end{center}

In analogy to the equivalence theorem of Section III, similar quantum flow algorithms can also be defined in the case of the Hermitian downfolding
(see Ref.\added{ }\onlinecite{kowalski2021dimensionality}). Although, due to non-commutative character of generators defining anti-Hermitian $\sigma_{\rm int}(\mathfrak{h}_i)$ and $\sigma_{\rm ext}(\mathfrak{h}_i)\;\;(i=1,\ldots,M)$, certain approximations has to be used 
(mainly associated with the use of the Trotter formula), similar flow can be defined for the Hermitian case (see Fig.\ref{fig1}). In this flow\added{,} we couple Hermitian eigenvalue problems 
corresponding to various active spaces (defined by sub-algebras $\mathfrak{h}_i$ and corresponding  effective Hamiltonians 
$H^{\rm eff}(\mathfrak{h}_i)\;\;(i=1,\ldots,M)$). 
The main advantage of this approach is the fact that larger  sub-spaces of the Hilbert space  can be sampled by a number of small-dimensionality active-space problems. This feature eliminates certain problems associated with (1) the need of using large qubits registers to represent the whole system, (2) qubit mappings of the basic operators, and (3) assuring anti-symmetry of the wave function of the entire system. For example, \deleted{the} problem (3) is \deleted{now} replaced by \deleted{the} procedures \replaced{that assure}{of the assuring} the anti-symmetry of the wave-functions of sub-systems defined by the active space generated by various $\mathfrak{h}_i$ ($D_i\ll N$ as shown in Fig.\ref{fig1}). This approach is \deleted{an} ideal for developing quantum algorithm\added{s} that 
take full advantage of the sparsity (or the local character of the correlation effects)  of the system and \replaced{uses}{using} only a small fraction\deleted{s} of  qubits ($D_i$, see Fig.\added{ }\ref{fig1}) needed to describe the system \replaced{represented}{described} by $N$ \replaced{spin orbitals}{spinorbitals}.

\begin{widetext}
\begin{center}
\begin{figure}
	\includegraphics[angle=0, width=0.78\textwidth]{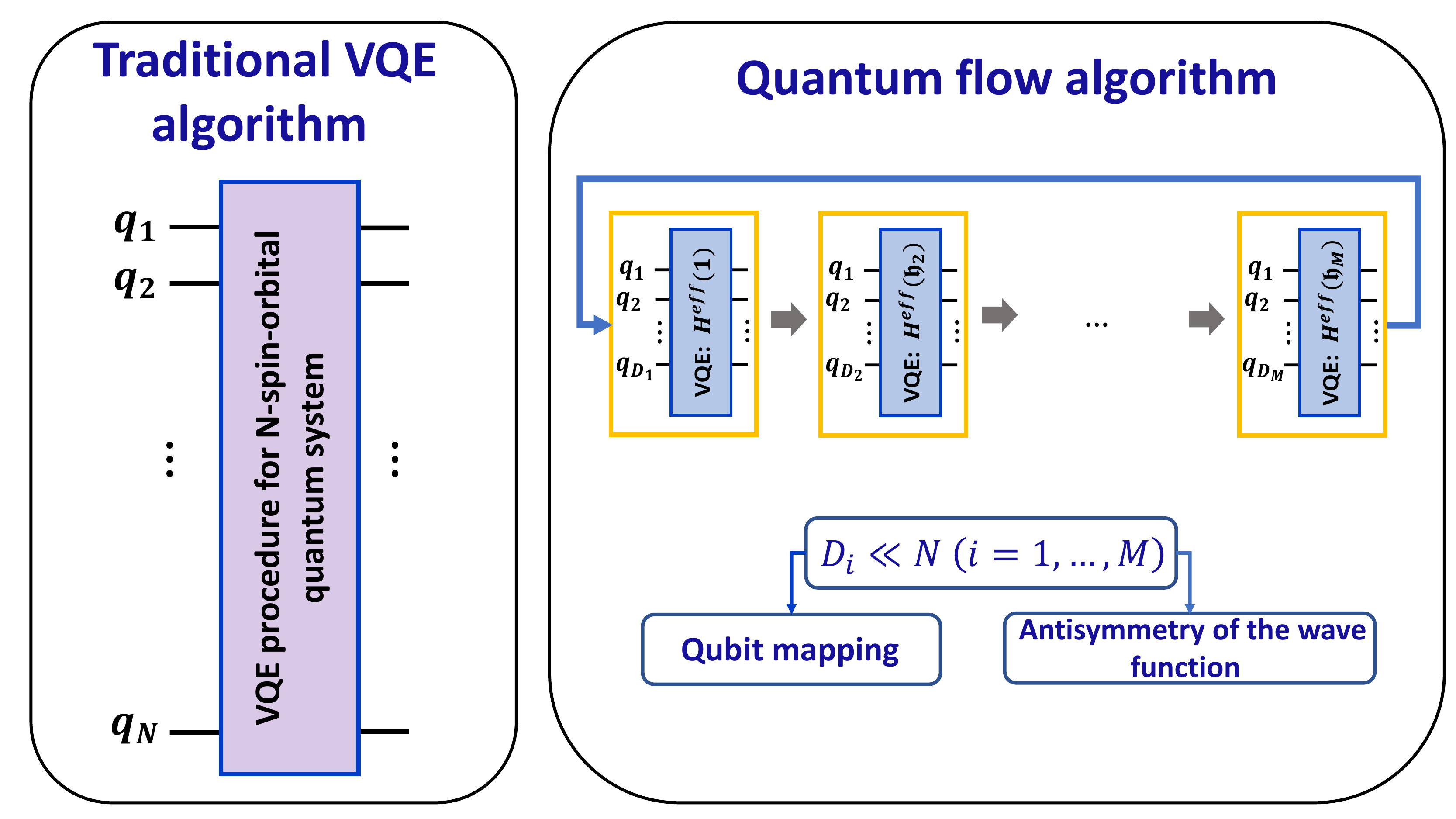}
	\caption{A schematic representation of the quantum flow algorithm. In this algorithm, the computational problem for a large space is translated into the computational flow involving coupled eigenvalue problems involving Hermitian 
	$H^{\rm eff}(\mathfrak{h}_i) \; (i=1,\ldots,M)$ 
	(see text for more details)}
\label{fig1}
\end{figure}
\end{center}
\end{widetext}


\section{Multi-component CC downfolding}

The development of computational algorithms for composite quantum systems keep\added{s} attracting a lot of attention in the field of quantum computing. \replaced{T}{A t}ypical examples are related to quantum electrodynamics, nuclear physics, and quantum chemistry. 
In quantum chemistry\added{,} this effort is related to the development of methods for non-perturbative coupling of electronic degrees of freedom with strong external fields \cite{haugland2020coupled,pavovsevic2021polaritonic} and formulations going beyond Born-Oppenheimer approximation.\cite{nakai2003many,ellis2016development,pavosevic2018multicomponent,pavosevic2021multicomponent}
Given the current status of quantum computing technology\added{,} it is important to provide techniques for compressing the dimensionality of these problems or finding an effective potential experienced by the one type of particles.

For simplicity, in this \replaced{s}{S}ection we will consider a fictitious system composed of two types of Fermions A and B, defined by two sets of creation/annihilation operators 
$\lbrace a_{\alpha}, a_{\alpha}^{\dagger} \rbrace_{\alpha=1}^{N_A}$ 
and
$\lbrace b_{\beta}, b_{\beta}^{\dagger} \rbrace_{\beta=1}^{N_B}$ (these operators should not be confused with the notation used in Section II)
satisfying typical Fermionic anti-commutation relations ($[.,.]_{+}$) 
and commuting ($[.,.]_{-}$) between themselves
\begin{eqnarray}
&& [a_{\alpha},b_{\beta}]_{-} = [a_{\alpha},b^{\dagger}_{\beta}]_{-} = 0 \;,\label{cot1} \\
&& [a^{\dagger}_{\alpha},b_{\beta}]_{-} = [a^{\dagger}_{\alpha},b^{\dagger}_{\beta}]_{-} = 0 \;.\label{cot2}
\end{eqnarray}
We will also assume specific form of the Hamiltonian
\begin{equation}
    H_{AB}=H_A+H_B+V_{AB}
\end{equation}
where $H_A$, $H_B$, \replaced{and}{a bd} $V_{AB}$ describe sub-systems A, B, and interactions between A and B, respectively. 
We will also assume that the interaction part commutes with the particle number operators $n_A$ and $n_B$ for systems A and B, i.e.,
\begin{eqnarray}
&& [V_{AB},n_A]_{-}=[V_{AB},n_B]_{-}=0 \;, \label{cot3} \\
&& n_A = \sum_{\alpha=1}^{N_A} a^{\dagger}_{\alpha} a_{\alpha} \;\;,\;\;
n_B = \sum_{\beta=1}^{N_B} b^{\dagger}_{\beta} b_{\beta} \;.
\end{eqnarray}
This situation\deleted{s} is typically encountered in models relevant to  non-Born-Oppenheimer approaches in electronic structure theory \cite{nakai2003many,ellis2016development,pavosevic2018multicomponent,pavosevic2021multicomponent} and nuclear structure theory.\cite{PhysRevC.69.054320}

Let us assume that the correlated ground-state wave function can be represented in the form of single reference CC wave function
\begin{equation}
    |\Psi_{AB}\rangle = e^{T_{AB}} |\Phi_{AB}\rangle  \;,
    \label{cccom}
\end{equation}
where cluster operator contains excitations correlating sub-system A ($T_A$) and sub-system B ($T_B$) as well as collective excitations involving both sub-systems ($S_{AB}$), .i.e.,
\begin{equation}
    T_{AB}=T_A+T_B+S_{AB} \;.
    \label{tab}
\end{equation}
\replaced{T}{where t}he reference function $|\Phi_{AB}\rangle$ is a reference function for the composite system which is assume to be represented as 
\begin{equation}
    |\Phi_{AB}\rangle = \Omega_A \Omega_B |0\rangle \;,
    \label{phiab}
\end{equation}
where $|0\rangle$ represents  physical vacuum\deleted{.} and $\Omega_A$ and $\Omega_B$
are string of $a^{\dagger}_{\alpha}$/$b^{\dagger}_{\beta}$ operators distributing electrons among occupied levels of sub-systems A and B, respectively. 

The energy-dependent CC equation for the composite system take\added{s} the form 
\begin{equation}
(P_{AB}+Q_{AB}) H_{AB} e^{T_{AB}} |\Phi_{AB}\rangle
= E_{AB} e^{T_{AB}} |\Phi_{AB}\rangle \;, 
\label{ccab} 
\end{equation}
  \\ 
where $E_{AB}$ is the energy of the composite system, $P_{AB}$ is a projection operator onto the reference function $|\Phi_{AB}\rangle$, and
projection operator $Q_{AB}$ can be decomposed as follows\replaced{:}{,}
\begin{equation}
 Q_{AB}=Q_A + Q_B + Z_{AB} \;,
 \label{zab}
\end{equation}
where \replaced{$Q_A$ and $Q_B$ are}{$Q_A$($Q_B$) is} the  projection operator\added{s} onto excited Slater determinants obtained by exciting particles within sub-system A \replaced{and B}{(B)} from $|\Phi_{AB}\rangle$\added{, respectively,} and 
$Z_{AB}$ corresponds to the projection operator onto sub-space spanned by excited Slater determinants where \replaced{fermion particles}{particles Fermions} of type A and B are excited simultaneously. 

By projecting \added{Eq.} (\ref{ccab}) onto $(P_{AB}+Q_A)$ and introducing the resolution of identity $e^{T_B+S_{AB}}e^{-T_B-S_{AB}}$ one obtains
\begin{equation}
(P_{AB}+Q_{A}) e^{T_B+S_{AB}}(\bar{H}_{AB, {\rm ext}}-E_{AB}) e^{T_{A}} |\Phi_{AB}\rangle
= 0 \;, 
\label{ccd1} 
\end{equation}
where
\begin{equation}
\bar{H}_{AB, {\rm ext}}=
e^{-T_B-S_{AB}} H_{AB} e^{T_B+S_{AB}} \;.
\label{ccd2}
\end{equation}
In analogy to analysis in \replaced{Ref. \onlinecite{safkk}}{Ref.\cite{safkk}} the role of $e^{T_B+S_{AB}}$ in Eq.\added{ }(\ref{ccd1})\added{,}  reduces to the unit operator. This is a consequence of the fact that the operator $T_B+S_{AB}$ produces excitations within sub-system B, which are subsequently eliminated by the $Q_A$ projection operator. Consequently, Eq.\added{ }(\ref{ccd1}) takes the form:
\begin{equation}
    H^{\rm eff}(A) e^{T_A} |\Phi_{AB}\rangle =E_{AB} e^{T_A}
    |\Phi_{AB}\rangle
    \;, \label{ccd3}
\end{equation}
where the downfolded/effective Hamiltonian 
$H^{\rm eff}(A)$
is defined as 
\begin{equation}
 H^{\rm eff}(A)=(P_{AB}+Q_A) \bar{H}_{AB, {\rm ext}}  (P_{AB}+Q_A)\;.
\label{ccd4}
\end{equation}
The above result  shows that once $T_B$ and $S_{AB}$ amplitudes are know (or can be effectively approximated) the energy of the entire system can be calculated performing simulations on the sub-system A using effective Hamiltonian $H^{\rm eff}(A)$.

In addition to the simplest downfolding procedure described above, there are several other possible scenarios how downfolding procedures can be defined for the composite system:
\begin{itemize}
 \item the utilization of second downfolding procedure to the $H^{\rm eff}(A)$ in reduced-size active space for sub-system A,
 \item the utilizaton of the composite  active space that is representd by tensor product of active spaces for sub-systems A and B. 
\end{itemize}
These techniques are especially interesting for the the explicit inclusion of nuclear degrees  of freedom  (for Fermionic nuclei) in the effective Hamiltonians describing electronic degrees of freedom in the non-Born-Oppenheimer formulations. 

A Hermitian extension of the downfolding procedure can be accomplished by utilizing DUCC Ansatz for the composite system given by the expansion
\begin{equation}
    |\Psi_{AB}\rangle = e^{\sigma_B+\rho_{AB}} e^{\sigma_A} |\Phi_{AB}\rangle \;,
    \label{duccab1}
\end{equation}
where $\sigma_A$, $\sigma_B$, and  $\rho_{AB}$ are the anti-Hermitian operators defined by the cluster amplitudes with indices belonging to  sub-systems A,\added{ }B, and amplitudes defined by a mixed indices involving basis functions on A and B, respectively. As in the non-Hermitian case of downfolding 
discussed in this Section, we will focus on the downfolding of the entire sub-system B into the effective Hamiltonians for sub-system A. 
Since creation/annihilation operators correspond to the sub-systems A and B, 
the exactness of the above expansion can be obtained as a generalization of the procedure based on the elementary Givens rotations discussed in \replaced{Ref. \onlinecite{evangelista2019exact}}{Ref.\cite{evangelista2019exact}}. For the specific case discussed in this Section (based on the downfolding of  the entire B sub-system) one should  assume that all basis functions defining sub-system A are defined as active indices
(see Ref.\cite{downfolding2020t}\note{what reference} for details). 

Substituting Eq.\added{ }(\ref{duccab1}) into the Schr\"odinger equations
and projecting onto $(P_{AB}+Q_A)$\added{,} one arrives the following form of the equations 
\begin{equation}
    H^{\rm eff}_{DUCC}(A)e^{\sigma_A}|\Phi_{AB}\rangle = E_{AB} e^{\sigma_A}|\Phi_{AB}\rangle \;,
    \label{duccab0}
\end{equation}
where
\begin{equation}
    H^{\rm eff}_{DUCC}(A)=
    (P_{AB}+Q_A) 
    \bar{H}_{AB,{\rm ext}}
    (P_{AB}+Q_A)  \;,
    \label{duccab2}
\end{equation}
and
\begin{equation}
    \bar{H}_{AB,{\rm ext}} = 
    e^{-\sigma_B-\rho_{AB}} H e^{\sigma_B+\rho_{AB}} \;.
    \label{duccab3}
\end{equation}
Again, the energy of the full system can be probed by sub-system A using effective Hamiltonian $H^{\rm eff}_{DUCC}(A)$. For example, one can envision the utilization of Eq.\added{ }(\ref{duccab0}) in the context of coupling nuclear and electronic degrees of freedom. In this case, sub-system A is represented by electrons while system B corresponds do nuclei obeying Fermi statistic. If $\sigma_B$ and $\rho_{AB}$ can be effectively approximated then the $H^{\rm eff}_{DUCC}(A)$ Hamiltonian describes the behavior of electron in the presence of "correlated" nuclei. The intensive development of the CC models beyond Born-Oppenheimer approximations 
\cite{nakai2003many,pavosevic2018multicomponent,pavosevic2021multicomponent} 
provides a reference for building approximate, for eaxmple, perturbative, form of $\sigma_B$ and $\rho_{AB}$ according formula analogous to Eqs.\added{ }(\ref{sext}), which requires the knowledge of $T_{B}$ and 
$S_{AB}$ 
to determine 
$\sigma_B$ and $\rho_{AB}$, respectively.

\section{Extraction of the analytical form of interactions in many-body systems}


In standard formulations of downfolding methods it is  assumed (see Refs.\added{ }\onlinecite{bauman2019downfolding,metcalf2020resource})) that downfolded Hamiltonians are 
dominated by one- and two-body effects, i.e., using the language of second quantization $H^{\rm eff}$ can be approximated as 
(for simplicity, let us assume that only virtual orbitals are downfolded)
\begin{equation}
 H^{\rm eff}_{DUCC} \simeq \sum_{PQ} \chi^P_Q a_Q^{\dagger} a_P + \frac{1}{2} \sum_{P,Q,R,S} \chi^{PQ}_{RS} a_R^{\dagger} a_S^{\dagger} a_Q a_P \;,
\label{gammaph}
\end{equation}
where $P,Q,R,S$ indices, $\chi^P_Q$, and $\chi^{PQ}_{RS}$ represent active \replaced{spin orbitals}{spinorbitals} and effective one- and two-body interactions, respectively
(non-antisymmetrized matrix elements $\chi^{PQ}_{RS}$ are employed in 
(\ref{gammaph})).
Once the set of  $\lbrace \chi^P_Q,\chi^{PQ}_{RS}\rbrace $ is  known (at the end of flow procedure)
this information can be further used to derive an analytical form of effective inter-electron interactions. This can be accomplished by 
fitting the general form of one-body $u$ and two-body $g$ interactions defined as  functions of to-be-optimized parameters
${\bm \gamma}$/${\bm \delta}$ as well as $r_1$, $r_2$, $r_{12}=|r_1-r_2|$, 
$\nabla_1$, $\nabla_2$, etc. operators:
\begin{eqnarray}
u &=& u({\bm \gamma},r_1,\nabla_1,\ldots) \;,
\label{f1} \\
g &=& g({\bm \delta},r_1,r_2,r_{12},\nabla_1,\nabla_2,\ldots) \;,
\label{f2}
\end{eqnarray}
These effective interactions replace standard one- and two-body interactions in non-relativistic quantum chemistry and are defined to minimize the 
discrepancies with $\lbrace \chi^P_Q,\chi^{PQ}_{RS}\rbrace $ for a given discrete molecular \replaced{spin-orbital}{spinorbital} set, i.e.,
\begin{eqnarray}
&\min_{\bm \gamma} \lbrace \sum_{PQ} |u^P_Q({\bm \gamma})-\chi^P_Q| \rbrace & \;,\label{minf}  \\
&\min_{\bm \delta}
\lbrace
\sum_{PQRS} |g^{PQ}_{RS}({\bm \delta})-\chi^{PQ}_{RS}|
\rbrace
& \;,
\;\label{ming}
\end{eqnarray}
where
\begin{widetext}
\begin{eqnarray}
u^P_Q({\bm \gamma}) &=& \int dx_1 \phi_Q(x_1)^{*}
u({\bm \gamma},r_1,\nabla_1,\ldots) \phi_P(x_1) \;,
\label{int1f} \\
g^{PQ}_{RS}({\bm \delta}) &=& \int dx_1 dx_2 \phi_R(x_1)^{*} 
\phi_S(x_2)^{*}
g({\bm \delta},r_1,r_2,r_{12},\nabla_1,\nabla_2,\ldots) \phi_P(x_1) \phi_Q(x_2)\;.
\label{int2g}
\end{eqnarray}
\end{widetext}
We believe that the utilization of efficient non-linear optimizers or machine learning techniques can provide an effective form of the interactions $u$ and $g$ defined in small-size active spaces. These effective interactions can be utilized in low-order methodologies, including  Hartree-Fock (HF)  and density functional theories (DFT). In the latter case 
functions $u$ and $g$ can be utilized to develop/verify new forms of exchange-correlations functionals. The access to the analytical form of the inter-electron interactions can also enable affordable and reliable ab-initio dynamics driven by low-order methods.

\section{Conclusions}

In this paper\added{,} we briefly review \added{the} current state of  two variants of CC downfolding techniques.
While the non-Hermitian downfolding and resulting active-space Hamiltonians are not a primary target for quantum computing, the equivalence theorem opens new \replaced{possibilities}{possibility} regarding forming systematic  reduced-scaling frameworks based on the quantum flow equations. In contrast to the existing reduced scaling CC formulations, where the notion of electron pair is rather descriptive and is based on the partitioning of the correlation energy with respect to contributions that can be indexed by pairs of the occupied orbitals, the present formalism defines the pair through the corresponding effective Hamiltonian. This fact has \added{a} fundamental advantage over {\it ad hoc} localization procedures - it allows in a natural way to introduce the pair density matrix. It also \replaced{allows}{open the way} for \added{a} more systematic way of introducing certain classes of higher-rank excitations. The double unitary CC Ansatz provides a natural many-body language to introduce Hermitian downfolded  representation of many-body Hamiltonians in reduced-dimensionality active spaces. To approximate non-terminating commutator expansion of downfolded Hamiltonians\added{,} we use finite  commutator expansions. On the LiF example, we demonstrated that the inclusion of double commutator terms leads to systematic improvements of the results obtained with single commutator expansion even in \added{a} situation when \added{an} active space is not providing a good zero-th order approximation of correlation effects. It should be also stressed that the downfolded \replaced{Hamiltonians}{Hamitlonians} based \replaced{on the}{ont he} double commutator expansion are capable of reducing the error of energies obtained by the diagonalization  of \added{the} bare-Hamiltonian  in \added{the same-size} active space by more than an order of magnitude (in fact\added{,} for the 1.0R$_e$ and 5.0R$_e$ one could witness 30- and 60-fold reduction in energy errors with respect to accurate CC results obtained by correlating all molecular orbitals). 

In the second part of the paper\added{,} we extended  non-Hermitian and Hermitian downfolding to multi-component quantum systems. As an example, we used the model system composed of two types of Fermions, which epitomize typical situations encountered in nuclear physics and for certain types of nuclei in non-Born-Oppenheimer electronic structure theory. We have also outlined an approximate procedure to extract the semi-analytical form of the one- and two-body inter-electron interactions in active space based on the minimization procedure 
utilizing one- and two-body interactions defining downfolded Hamiltonians. In the future\added{,} we will explore the usefulness of \replaced{machine learning}{mchine learinign} techniques for this procedure.

\section{Acknowledgement}
This work was supported by the "Embedding QC into Many-body Frameworks for Strongly Correlated  Molecular and Materials Systems'' project, which is funded by the U.S. Department of Energy, Office of Science, Office of Basic Energy Sciences (BES), the Division of Chemical Sciences, Geosciences, and Biosciences.
Part of this work was supported by  the Quantum Science Center (QSC), a National Quantum Information Science Research Center of the U.S. Department of Energy (DOE).



%

\end{document}